\documentclass[twocolumn, amsfonts, amssymb, aps, nofootinbib, noeprint]{revtex4-2}
\usepackage[T1]{fontenc}
\usepackage{tgtermes}
\usepackage{amsmath}
\usepackage{empheq}
\usepackage{enumitem}
\usepackage{graphicx}
\usepackage{svg}
\usepackage{booktabs}
\usepackage[braket, qm]{qcircuit}
\usepackage{braket}
\usepackage{hyperref}
\usepackage{placeins}
\usepackage{xcolor}
\usepackage{soul}
\usepackage{tikz}
\usepackage{bbm}
\usepackage{subcaption}
\hypersetup{
	colorlinks   = true, 
	urlcolor     = blue, 
	linkcolor    = blue, 
	citecolor   = red 
}
\usepackage[normalem]{ulem}

\newcommand{\CZ}{\textsf{CZ }}

\newcommand{\HGate}{\textsf{H }}
\newcommand{\SGate}{\textsf{S }}

\newcommand{\CP}{Clifford+Pauli}
\begin{document}

\title{Fourier expansion in variational quantum algorithms}

\author{Nikita A. Nemkov}\email{nnemkov@gmail.com}
\affiliation{Russian Quantum Center, Skolkovo, Moscow 143026, Russia}
\affiliation{National University of Science and Technology ``MISIS”, Moscow 119049, Russia}
\author{Evgeniy O. Kiktenko}
\affiliation{Russian Quantum Center, Skolkovo, Moscow 143026, Russia}
\affiliation{National University of Science and Technology ``MISIS”, Moscow 119049, Russia}
\author{Aleksey K. Fedorov}\email{akf@rqc.ru}
\affiliation{Russian Quantum Center, Skolkovo, Moscow 143026, Russia}
\affiliation{National University of Science and Technology ``MISIS”, Moscow 119049, Russia}

\begin{abstract}
The Fourier expansion of the loss function in variational quantum algorithms (VQA) contains a wealth of information, yet is generally hard to access. We focus on the class of variational circuits, where constant gates are Clifford gates and parameterized gates are generated by Pauli operators, which covers most practical cases while allowing much control thanks to the properties of stabilizer circuits. We give a classical algorithm that, for an $N$-qubit circuit and a single Pauli observable, computes coefficients of all trigonometric monomials up to a degree $m$ in time bounded by $\mathcal{O}(N2^m)$. Using the general structure and implementation of the algorithm we reveal several novel aspects of Fourier expansions in \CP{} VQA such as (i) reformulating the problem of computing the Fourier series as an instance of multivariate boolean quadratic system (ii) showing that the approximation given by a truncated Fourier expansion can be quantified by the $L^2$ norm and evaluated dynamically (iii) tendency of Fourier series to be rather sparse and Fourier coefficients to cluster together (iv) possibility to compute the full Fourier series for circuits of non-trivial sizes, featuring tens to hundreds of qubits and parametric gates.

\end{abstract}

\maketitle

\tableofcontents

\section{Introduction and results}
Variational quantum algorithms (VQA)~\cite{Cerezo2021} are the leading candidates to make the most out of current NISQ devices~\cite{Preskill2018, Bharti2021}. While the scope of potential VQA applications is extremely broad, there are also many theoretical and practical limitations. VQA are hybrid algorithms, using classical optimization to train parameterized quantum circuits, and in this sense they are similar to the classical machine learning models. The loss function of a VQA is defined as an average of some observable in the state, prepared by the parameterized quantum circuit. The structure of VQA loss landscape is of central importance, because the efficiency of the classical optimization largely determines the quality of solutions obtained by VQA. Accessible shapes of the loss function also determine the expressive power of the quantum machine learning models~\cite{Biamonte2017}.

Typically, VQA are trained by gradient-based methods and their local properties are of the most interest. At the same time, the structure of parametrized quantum circuits makes Fourier series representation a natural and rich language for description of VQA loss functions. We now briefly survey some of the relevant results. Under the assumption that generators of parametric gates have commensurable eigenvalues, the Fourier series in fact truncates to a trigonometric polynomial. As shown in Ref.~\cite{Schuld2020}, accessible frequencies in this expansion are determined by the spectrum of the generators, while the coefficients depend on the structure of the circuit and the observable. In Refs.~\cite{Schuld2020, GilVidal2020} this observation was used to highlight the importance of data encoding in quantum machine learning models. Recently, there has been an interest in quantitatively studying the expressive power of machine learning models based on the properties of their Fourier expansion~\cite{Atchade-Adelomou2023, Casas2023}.
In Refs.~\cite{Schreiber2022, Landman2023} the Fourier representation was used to de-quantize a class of quantum machine learning models. Paper~\cite{Fontana2022} investigates the case when the Fourier series is rather sparse, so that the loss landscape can be efficiently recovered with limited experimental data. An interesting proposal made in Ref.~\cite{Fontana2022a} shows that noise in VQA can be detected by observing inaccessible frequencies and mitigated by filtering them out. We also note that the fundamental result on the NP-hardness of training general VQA~\cite{Bittel2021} relies on their loss functions being trigonometric polynomials.

While the Fourier representation can be a very convenient tool to characterize variational loss functions, it has its limitations. When generators of the parametric gates square to identity, which is the most common case, the Fourier series for a VQA with $M$ parameters is a multivariate trigonometric polynomial containing up to $3^M$ terms. Exponential growth of accessible terms makes the Fourier series an impractical description, unless the actual distribution of coefficients is very sparse (e.g. the number of non-zero terms only grows polynomially with $M$). Computing the Fourier coefficients is also a challenge. For example, evaluating the lowest order, constant Fourier term, amounts to finding the loss function averaged over all parameter configurations, and there seems to be no efficient recipe for that in general.  

In this paper, we restrict attention to a special class of parametrized quantum circuits, which we refer to as \textit{\CP{}} circuits. Parametric gates in \CP{} circuits are exponentials of Pauli strings, while constant gates are Clifford gates. This is in fact a very large class of circuits that includes the majority of most studied VQA, such Quantum Approximate Optimization Algorithm (QAOA) \cite{Farhi2014}, Hardware-Efficient Ansatz (HEA), \cite{Kandala2017}, Unitary Coupled Cluster Ansatz \cite{Romero2018}. Special properties of stabilizer circuits \cite{Gottesman1997, Aaronson2004} give an essential technical leverage to study the Fourier expansion of \CP{} circuits in quantitative details. 

The interplay between properties of stabilizer circuits and VQA have been explored previously, mainly in the context of quantum chemistry~\cite{Cao2019, Bauer2020, Fedorov2022VQE}. In particular, initialization methods based on perturbative expansion around Clifford points~\cite{Mitarai2022} or on the discrete search through the space of Clifford circuits~\cite{Cheng2022, Ravi2022} have been developed, ansatz structures~\cite{Ryabinkin2020, Brown2022a} and partitioning schemes~\cite{Schleich2023}, based on the properties of Clifford gates have been proposed. In this work, however, we focus on a different scope of questions.

Our core technical contribution is an efficient classical algorithm computing all Fourier coefficients in the loss function up to level $m$, with time complexity bounded by $O(N2^m)$, where $N$ is the number of qubits. Note that in general, the complexity of the Fourier series is not limited directly by the number of qubits $N$, but rather by the total number of parametric gates $M$. For \CP{} VQA this observation thus admits a concrete realization.

The algorithm has both theoretical and practical utility. On the theoretical side, we show that typical Fourier series are much sparser than anticipated in the general case. For \CP{} circuits with a single Pauli observable, the number of coefficients is upper bounded by $2^M$, and for the worst case behavior expected in practice we find $\left(\frac32\right)^{M}$. We also show, that truncating the Fourier polynomial below the maximum degree gives an approximation that can be quantified by the $L^2$ norm of the loss function and evaluated dynamically, as the algorithm proceeds. The number of terms contributing non-negligibly the functional norm is typically an exponential fraction of all terms, yet still growing exponentially itself.

On the practical side, we perform several case studies to probe the structure of Fourier expansions in more detail. In all examples we find that Fourier terms tend to cluster around some mean level, which is exactly $\frac{M}{2}$ for non-local random circuits, but much smaller for the local circuits with special structure, such as QAOA or HEA, making their Fourier expansion much sparser and easier to compute. 

Our algorithm is based on a simple recursive expansion of the loss function. A similar approach was described in the context of the Qubit Coupled Cluster Method~\cite{Ryabinkin2020}, and in the context of QAOA in Ref.~\cite{Hadfield2018}. However, there are some important distinctions with the previous work. First, we identify \CP{} circuits as the class to which the method is universally applicable, and treat the problem in general terms, as well as establish its direct relation to the Fourier series expansion. More importantly, we demonstrate how to significantly reduce the algorithm cost by pruning some branches of the recursive expansion early, based on filtering by expectation values. While this does not change the large $M$ asymptotic, for practical cases the difference is essential. For example, for random \CP{} circuits on $N=50$ qubits, it allows increasing the depth of circuits that can be handled from $M=30$ to $M=80$ without changing the computational budget. Finally, we formulate the problem of computing all non-zero coefficients in the Fourier expansion as an instance of the multivariate boolean quadratic problem, which allows us to argue that our algorithm is likely not far from optimal yet points towards potential improvements.

\section{Parameterized quantum circuits}\label{sec VQA}

In this section, we establish some notation, describe basic properties of variational circuits and their loss functions, and discuss how Fourier expansion arises in this context.
\subsection{Trigonometric expansion of the unitary}
We will assume that a unitary matrix $U(\pmb{\phi})$ of a parameterized quantum circuit takes the following form: 
\begin{align}
	U(\pmb{\phi})=C_M P_M(\phi_M)\dots C_2P_2(\phi_2)C_1 P_1(\phi_1) C_0 \ . \label{U phi}
\end{align}
Here $C_m$ are constant gates, $P_m(\phi)=e^{-\imath\frac{\phi}{2} G_m}$ are single-parameter rotations, and $\pmb{\phi}=(\phi_1,\ldots,\phi_M)$ is a vector of parameters.
We will assume that Hermitian generators of the parameterized gates square to identity $G_m^2=\mathbbm{1}$, so that
\begin{align}
	P_m(\phi)=\mathbbm{1}\cos\frac{\phi}{2}-\imath G_m\sin\frac{\phi}{2} \ .
\end{align}
Applying this relation to each parametric gate in the circuit one obtains the following formal trigonometric expansion containing $2^M$ terms
\begin{align}
	U(\pmb{\phi})=\sum_{I\in\{0,1\}^M} U_I t_I\left(\frac{\pmb{\phi}}2\right) \ . \label{U expansion}
\end{align}
Here $I=(I_1,\ldots,I_M)$ with $I_m\in\{0,1\}$ 
is a multi-index, $t_I(\pmb{\phi})$ is a multivariate trigonometric monomial of order $M$
\begin{align}
	t_I(\pmb{\phi})=\prod_{m=1}^M t_{I_m}(\phi_m) \label{t def} \ ,
\end{align}
where each term in the product is defined by
\begin{align}
	t_i(\phi)=\cos^{1-i}\phi\sin^i\phi=\begin{cases}\cos\phi,\quad i=0\\\sin\phi,\quad i=1\end{cases} \ .
\end{align}
We note that coefficient matrices $U_I$ correspond to the circuit unitary, evaluated at specific values 
\begin{equation}
    U_I=U(\pmb{\phi}=\pi I)=\alpha C_MG_M^{I_M}\ldots C_1G_M^{I_1}C_0 \ ,
\end{equation}
where $\alpha$ is a phase factor $\alpha=(-\imath)^{\sum_m I_m}$.

\subsection{Fourier expansion of the loss function}
The loss function $F(\pmb\phi)$ of a variational algorithm is defined by the average of some Hermitian operator $H$, often referred to as \textit{the Hamiltonian}, in the state prepared by the circuit 
\begin{align}
 F(\pmb{\phi})=\Braket{0|U^\dagger(\pmb\phi)HU(\pmb\phi)|0} \ . \label{F def}
\end{align}
Here and in the following $\ket{0}=\ket{0}^{\otimes N}$ is the all zeros state of $N$ qubits. Substituting expansion \eqref{U expansion} into the loss function gives
\begin{align}
	F(\pmb{\phi})=\sum_{IJ}t_I\left(\frac{\pmb \phi}{2}\right)t_J\left(\frac{\pmb \phi}{2}\right)\Braket{0|U^\dagger_IHU_J|0} \ . \label{F exp}
\end{align}
In contrast to the expansion of unitary \eqref{U expansion}, which is homogeneous, expansion of the loss function contains trigonometric monomials of various degrees, see App.~\ref{app explicit} for details. Let us organize the Fourier expansion of the loss function by level
\begin{align}
	F(\pmb\phi)=\sum_{m=0}^M F_m(\pmb\phi) \ . \label{Fm def}
\end{align}
Each level $F_m(\pmb\phi)$ only involves monomials of order $m$.
There are $\binom{M}{m}$ possible parameter subsets at level $m$ each giving rise to $2^m$ trigonometric monomials. Hence, the total number of independent coefficients in the Fourier expansion is
\begin{align}
	\sum_{m=0}^M 2^{m}\binom{M}{m}=3^M \label{3M coeffs} \ .
\end{align}
\subsection{Fourier terms from averages}
Computing the Fourier series for generic loss functions appears to be a formidable task. Indeed, let us note that the constant term $F_0$ in the Fourier expansion can be thought of as the loss function,  averaged over all parameters
\begin{align}
	F_0=\Braket{F(\pmb\phi)}_{\pmb\phi}:=\frac{1}{(2\pi)^M}\int_0^{2\pi}\prod_{m=1}^Md\phi_m\,\, F(\pmb\phi) \ . \label{F avg}
\end{align}
This relation holds, because all higher levels $F_{m>0}(\pmb\phi)$ in the Fourier series trivially vanish when averaged. Higher level terms can be obtained similarly, see App.~\ref{app explicit}.

The average in Eq.~\eqref{F avg} can be expressed in a succinct form using orthogonality of trigonometric monomials $t_I$ \eqref{t def}  
\begin{align}
	\Braket{t_I(\pmb\phi) t_J(\pmb\phi)}_{\pmb\phi}=2^{-M}\delta_{IJ} \ . \label{t product}
\end{align}
Hence, averaging \eqref{F exp} yields
\begin{align}
	F_0=\frac{1}{2^M}\sum_{I\in\{0,1\}^M}\Braket{0|U_I^\dagger H U_I|0} \ . \label{F0}
\end{align}

Evaluating this expression explicitly seems to be out of reach for generic circuits. In terms of a classical simulation, computing any single expectation value in Eq.~\eqref{F0} is difficult on its own for a sufficiently large number of qubits. Even when the averages can be computed efficiently, either classically or provided an access to a quantum computer, equation \eqref{F0} still requires summing $2^M$ terms, infeasible for any significant number of parameters $M$. As we show in the next section, for \CP{} quantum circuits evaluating $F_0$ and in fact any particular monomial in the Fourier expansion is classically efficient. 
\section{\CP{} variational circuits}
\subsection{Definition and properties}
Let us first establish some notation relevant for stabilizer circuits. A single-qubit Pauli operator is simply an $X, Y, Z$ Pauli matrix or an identity, possibly with a phase $\pm1, \pm \imath$. An $n$-qubit Pauli operator is a tensor product of $n$ arbitrary single-qubit Pauli operators. Any two Pauli operators   $P_1$ and $P_2$ either commute or anti-commute: $P_1P_2=\pm P_2P_1$. Clifford gates $C$ are operators that transform every Pauli gate $P$ into some Pauli gate $P'$: $C^\dagger P C=P'$. The group of Clifford gates can be generated by the Hadamard gate $\HGate$, $\SGate=\sqrt{Z}$, 
and controlled NOT gate. Circuits consisting only of the Clifford gates applied to the stabilizer states (of which $|0\rangle$ is an example) can be efficiently simulated classically due to the Gottesman-Knill theorem~\cite{Aaronson2004}.

We define \CP{} variational circuits as a subset of variational circuits~\eqref{U phi}, where generators of parametric gates are Pauli operators and all constant gates are Clifford gates. For clarity of exposition, we also assume that the Hamiltonian $H$ is a Pauli operator. The case that is most relevant in practice, when the Hamiltonian is a polynomial-sized sum of Pauli operators, can be handled by linearity. We stress that both the Pauli generators and the Hamiltonian are allowed to have arbitrary weight, i.e. be supported on any number of qubits.
Note that Pauli rotations with generic angles are not Clifford gates, and hence \CP{} circuits can not be simulated efficiently using the stabilizer formalism. 

\CP{} circuits admit a simple canonical form, where all Clifford gates are eliminated. First, one uses commutation properties of Clifford and Pauli operators to drag all the Clifford gates to the very end of the circuit. Generators of Pauli rotations will generally change during the process. The Clifford gate $C$ accumulated at the end of the circuit is absorbed into the Hamiltonian $H\to C^\dagger H C$, which remains a Pauli operator. Hence, without loss of generality, we will assume that \CP{} variational circuit takes the following \textit{Pauli form}
\begin{align}
	U(\pmb\phi)=P_M(\phi_M)P_{M-1}(\phi_{M-1})\dots P_1(\phi_1) \ ,
\end{align}
where $P(\phi)=e^{-\imath\frac{\phi}{2}P}$ for some Pauli string $P$. We will use notation $(P_1\dots P_M|H)$ for a \CP{} circuit in Pauli form. Note that this form is not unique, as Pauli rotations with commuting generators can be swapped.

\subsection{Computing averages}
Now let us revisit the computation of the average loss function \eqref{F0}. Eq.~\eqref{F0} now takes the form
\begin{align}
    F_0=\frac{1}{2^M}\sum_{I\in\{0,1\}^M}\Braket{0|P_1^{I_1}\dots P_M^{I_M}HP_M^{I_M}\dots P_1^{I_1}|0} \ . \label{F0P}
\end{align}
We claim that this sum vanishes unless $H$ commutes with every $P_i$. Indeed, suppose there is some $P_m$ s.t. $P_mH=-HP_m$. Then, it is straightforward to see, that any two terms in the sum that only differ in the value of $I_m$ give opposite contributions. Hence, for the circuit $(P_1\dots P_M|H)$ we find
\begin{align}
	F_0=\begin{cases} \braket{0|H|0},\quad [H, P_i] = 0\quad \forall i\\ 0,\qquad\qquad \text{otherwise}\end{cases} \ .
\end{align}
Therefore, computing the average loss function for \CP{} circuits is a trivial task for any number of qubits and any number of parameters. In fact, as we show in the following, this applies to every individual term in the Fourier series. The difficulty of computing the full Fourier expansion then stems solely from the fact that the total number of non-vanishing coefficients can be exponentially large. In the next section, we present an efficient classical algorithm to compute Fourier expansion level by level.
\section{Classical algorithm}
\subsection{Expansion of the dressed Hamiltonian}
\begin{figure}
	\includegraphics[width=0.5\textwidth]{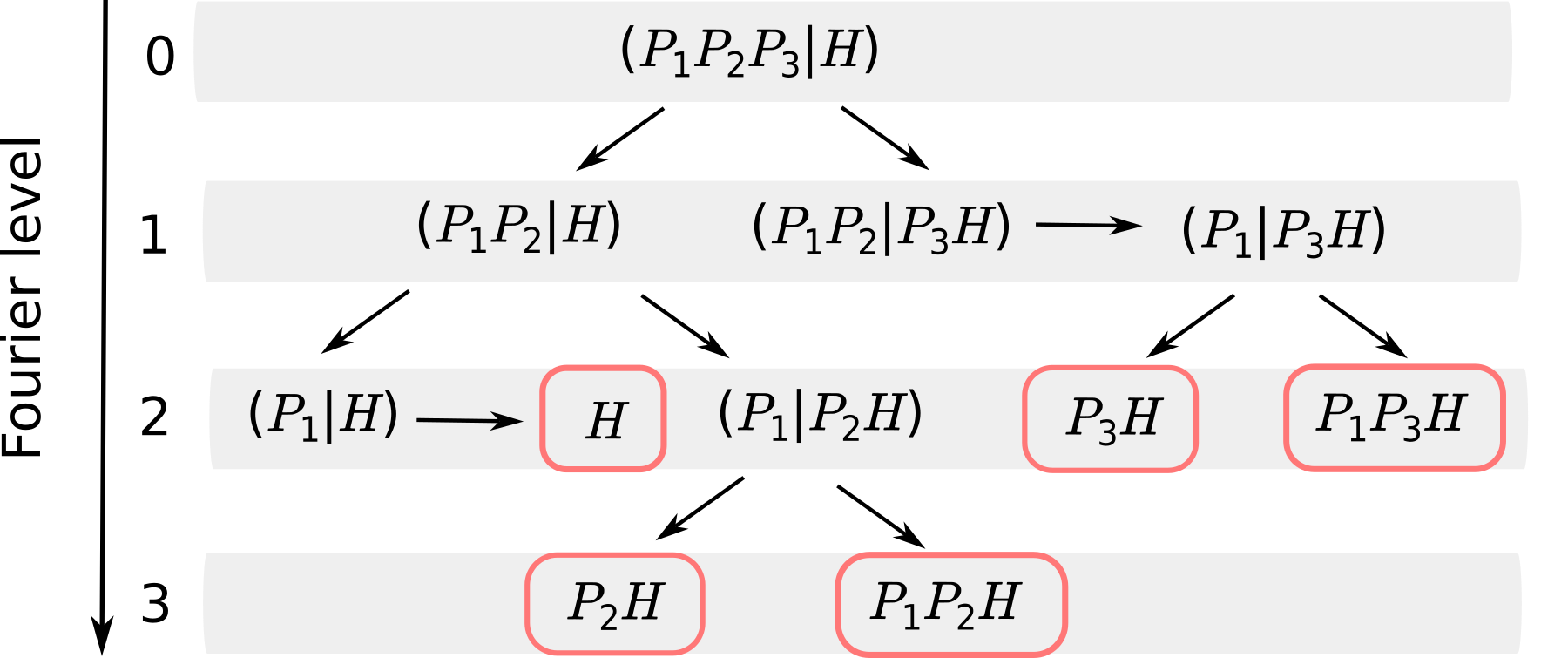}
	\caption{A sample diagram representing recursive expansion of a dressed Hamiltonian. Here $\{P_3,H\}=\{P_2,H\}=\{P_1,P_3H\}=\{P_1,P_2H\}=[P_2,P_3H]=[P_1,H]=0$.
 }
	\label{fig tree}
\end{figure}

Introduce the following notation for an operator conjugated by circuit's unitary
\begin{align}
	O[\pmb\phi]=U^\dagger(\pmb\phi)OU(\pmb\phi)\ .
    \label{eq:conj-def}
\end{align}
Following the quantum chemistry literature, we will call $H(\pmb\phi)$ the \textit{dressed Hamiltonian}. The loss function is the average of the dressed Hamiltonian in the all-zero state
\begin{align}
	 F(\pmb\phi)=\Braket{0|H[\pmb\phi]|0} \ . \label{F from H dressed}
\end{align}
Next, we make the following simple observation: for an arbitrary Pauli string $O$ it holds
\begin{align}
    P(\phi)^\dagger O P(\phi)=\begin{cases} O,\quad &[P, O]=0\\ 
    O\cos\phi+\imath PO\sin\phi,\quad &\{P,O\}=0\end{cases} \ ,
    \label{eq:simple-rule}
\end{align}
i.e. when the conjugating Pauli rotation $P(\phi)$ commutes with $O$, it cancels out, while for an anti-commuting Pauli rotation the result can be written as a sum of two Pauli operators. This gives a recurrence procedure to expand the dressed Hamiltonian. Indeed, it follows from \eqref{eq:simple-rule} that for an arbitrary Pauli string $O$  
\begin{align}
O[\pmb\phi^{(m)}]=
\begin{cases}
O[\pmb\phi^{(m-1)}], &[O, P_m]=0; \\
\begin{aligned}
    &O[\pmb\phi^{(m-1)}]\cos\phi_m +\\
    &~~+\imath (P_m O)[\pmb\phi^{(m-1)}]\sin\phi_m,
\end{aligned} 
    &\{O, P_m\}=0;
\end{cases} \label{eq:rec-rule}    
\end{align} 
Here $\pmb\phi^{(m)}$ the subset of the first $m\le M$ parameters $\pmb\phi^{(m)}:=(\phi_1,\ldots,\phi_m)$ and $O[\pmb\phi^{(m)}]$ is defined as in \eqref{eq:conj-def} with the conjugating unitary $U(\pmb \phi^{(m)}):=P_{m}(\phi_{m})\ldots P_1(\phi_1)$.

Repeatedly applying \eqref{eq:rec-rule} to $H[\pmb\phi]\equiv H[\pmb\phi^{(M)}]$ represents the dressed Hamiltonian as a sum of Pauli strings multiplied by trigonometric monomials, i.e. gives the Fourier expansion of the dressed Hamiltonian with operator coefficients. The recurrent expansion can be conveniently visualized as a binary tree, see Fig.~\ref{fig tree} for an example. The tree is constructed as follows. The nodes, which we refer to as the \textit{computational nodes}, correspond to variational circuits, specified by a list of Pauli generators and an observable $(P_1\dots P_m|O)$. If the observable $O$ anti-commutes with the last Pauli generator $P_m$, the node branches into two $(P_1\dots P_m|O)\to\cos\phi_m (P_1\dots P_{m-1}|O)+\imath\sin\phi_m (P_1\dots P_{m-1}|P_m O)$. For brevity, we omit coefficients at the diagram. Branching increases the Fourier level by one. If the last Pauli generator instead commutes with $O$, it is simply removed $(P_1\dots P_m|O)\to (P_1\dots P_{m-1}|O)$ and the Fourier level remains unchanged. We depict this by horizontal arrows at the diagram. When there are no Pauli generators left, the node contains the final observable encoding a single operator coefficient in the Fourier expansion of the dressed Hamiltonian.

The graphical representation makes several distinctive features of Fourier series for \CP{} circuits manifest. Let $n(m)$ be the number of resulting Fourier modes at level $m$. Introduce
\begin{align}
	\delta(m)=2^{-m}n(m),\quad \Delta(m)=\sum_{k=0}^m \delta(k) \label{delta def} \ .
\end{align}
For any \CP{} circuit and any Pauli Hamiltonian it holds
\begin{align}
    \Delta(M)=\sum_{m=0}^M 2^{-m}n(m)=1 \ , \label{invariant}
\end{align}
i.e. the weighted sum of populations at all Fourier levels is an invariant. 

This implies certain constraints on the distribution of Fourier terms. For example, the maximum number of Fourier terms $\sum_m n(m)$ is upper bounded by $2^M$ (when the last level is fully populated), cf. the bound for generic circuits $3^M$ \eqref{3M coeffs}. Importantly, the presence of any single Fourier term at level $m<M$ reduces the maximum possible amount of terms at other levels. For instance, if $n(0)=1$, i.e. $F_0\neq0$, all other Fourier terms vanish. 

For $N$-qubit circuits, processing each computational node only involves multiplying Pauli strings of length $N$, and hence has complexity $\mathcal{O}(N)$. Since the total number of nodes is at most $2^M$, the time complexity of the algorithm can be bounded by $\mathcal{O}(N2^M)$. In the following, we will give more detailed estimates for the expected number of computational nodes and complexity of the algorithm.

\subsection{Accounting for the expectation values} \label{sec pruning}
So far, we discussed the expansion of the dressed Hamiltonian. In turn, the loss function  is given by its expectation value in the all-zero state \eqref{F from H dressed}. A likely scenario is that the majority of the final Pauli observables have vanishing expectations, and hence do not contribute to the loss function. This observation allows significantly increasing the efficiency of the computation by pruning unfit branches in advance.

Let $\mathbb{F}_2^n$ be a vector space of binary strings of length $n$. For $k=(k_1,\dots,k_n)\in \mathbb{F}_2^n$ denote $\pmb Z(k) = \bigotimes_{i=1}^n Z^{k_i}$ ($\pmb X(k)$ is defined similarly). 
For any $n$-qubit Pauli operator $P$ one can define two vectors $P_{Z}, P_{X}\in\mathbb{F}_2^n$ such that
\begin{align}
    P=\alpha\,\, \pmb  Z(P_{Z}) \pmb X(P_{X}) \ ,
\end{align}
where $\alpha$ is a phase factor. One can think of $P_{Z}$ and  $P_{X}$ as coordinates of $P$ in the basis of Pauli $Z$ and $X$ strings, respectively. With the notation in place, we can explain how expectation values can be taken into account during the expansion of the dressed Hamiltonian.

A Pauli string $P$ has a non-zero expectation value $\Braket{0|P|0}\neq0$ iff $P_{X}=(0,\ldots,0)$. First assume, for simplicity, that $X$-vectors of the first $N$ Pauli generators $(P_{1})_{X},\ldots, (P_{N})_{X}$ are linearly independent and constitute a basis in $\mathbb{F}_2^N$. This implies that for every $O$, there is a unique vector $k\in \mathbb{F}_2^N$ such that 
\begin{equation}
    O_{X} = k_1(P_{1})_{X}+\ldots +k_N(P_{N})_{X} \ . \label{PX}
\end{equation}
Therefore, among all $2^N$ possible observables of the form $P_1^{t_1}\ldots P_N^{t_N}O$, that can be produced during recursive expansion of $(P_1\dots P_N|O)$, only a single one with $t_i=k_i$ can yield a non-zero expectation value (all other terms will have a non-zero $X$-component). Thus, instead of generating the full recursive expansion of $(P_1\dots P_N|O)$, which can contain up to $2^N$ nodes, we can first find $k$ from \eqref{PX}, and then check if this operator actually appears in the dressed Hamiltonian, i.e. it is compatible with the branching rules. This yields an exponential saving for large $N$.

Now let us lift the restriction of the first $N$ Pauli generators forming the basis. The necessary condition for $(P_1\ldots P_m|O)$ to have a non-zero expectation is that $O_{X}$ is contained in the span of 
$(P_1)_{X},\ldots,(P_m)_{X}$. Therefore, for each newly generated computational node, we can test if this condition is satisfied. If it is not, all final observables stemming from the expansion of this node have zero expectation values, and 
the node can be disregarded. 

As discussed in App.\ref{app pauli}, there is some room for further optimization based on the freedom to permute commuting Pauli generators. Also, at this point, we would like to spell out explicitly an elementary observation about the Fourier series of \CP{} circuits. For a single Pauli observable, all non-zero coefficients are given by averages of Pauli strings, and hence equal to $\pm1$.

\subsection{Truncated Fourier series as an approximation} \label{sec truncation}

Having many terms at low Fourier levels appears to be convenient, because this partially reduces a proliferation of coefficients at subsequent levels. This is further reinforced by the observation that each individual term at a lower level contributes exponentially more to the loss function than a term at a higher level. Intuitively, this is because the average absolute value of a trigonometric monomial of order $m$ is $(\pi/2)^{-m}$
and decays exponentially with degree $m$. At the same time, there can be exponentially more terms at higher levels. We can quantify this trade-off by evaluating the $L^2$ norm of the loss function. From orthogonality of trigonometric monomials \eqref{t product} it follows
\begin{align}
	   ||F||^2:=\Braket{|F(\pmb\phi)|^2}_{\pmb\phi}=\sum_{m=0}^M 2^{-m}l(m) \ ,
\end{align}
where $l(m)$ is the number of non-zero Fourier terms in the expansion of the loss function at level $m$, which is upper bounded by the number of non-zero terms in the expansion of the dressed Hamiltonian ($l(m)\le n(m)$).
Note that $n(m)-l(m)$ is the number of operators in the dressed Hamiltonian at level $m$ with zero expectation value (i.e. with non-trivial $X$ coordinate). Using \eqref{invariant}, we can then bound the $L^2$ norm of the loss function:
\begin{align}
	||F||^2\le 1 \ .
\end{align}
Let $F^{(m)}(\pmb\phi)$ denote the Fourier series truncated to the first $m$ levels. Then,
\begin{align}
	||F^{(m)}-F||^2\le \sum_{k=m+1}^M 2^{-k}l(k)\le 1-\Delta(m) \ . \label{F bound}
\end{align}
If $\Delta(m)$ is close to 1, i.e. sufficiently many terms are concentrated up to level $m$, the truncated Fourier series $F^{(m)}(\pmb\phi)$ gives a good approximation to the full loss function. Note that our recursive expansion generates $F^{(m)}(\pmb\phi)$ level by level, so the quality of the approximation can be gauged dynamically, and the computation stopped when the necessary accuracy is reached.

We need to mention two caveats related to the approximation result stated. First, while closeness in $L^2$ norm guarantees good approximation for most parameter configurations, it does not translate directly into point-wise convergence. Indeed, while the average absolute value of higher-level monomials \eqref{t def} is exponentially suppressed, their maximum values are independent of the order ($\operatorname{max}_{\pmb\phi}|{t}_{I}(\pmb\phi)|=1$). Second, the bound \eqref{F bound} may be too weak in practice, as it effectively assumes that all the final observables in the dressed Hamiltonian expansion above level $m$ have non-zero expectation values. In practice, we expect that only an exponentially small fraction of observables contributes to the loss function norm. Properly taking this into account can significantly strengthen the bound, but requires accounting for the structure of a particular circuit at hand. We illustrate this in a random circuit model discussed in Sec.~\ref{sec cases}.

\subsection{Is there a more efficient algorithm?} \label{sec complexity}
\begin{figure}
        \begin{subfigure}{0.4\textwidth}
        \includegraphics[width=0.75\hsize]{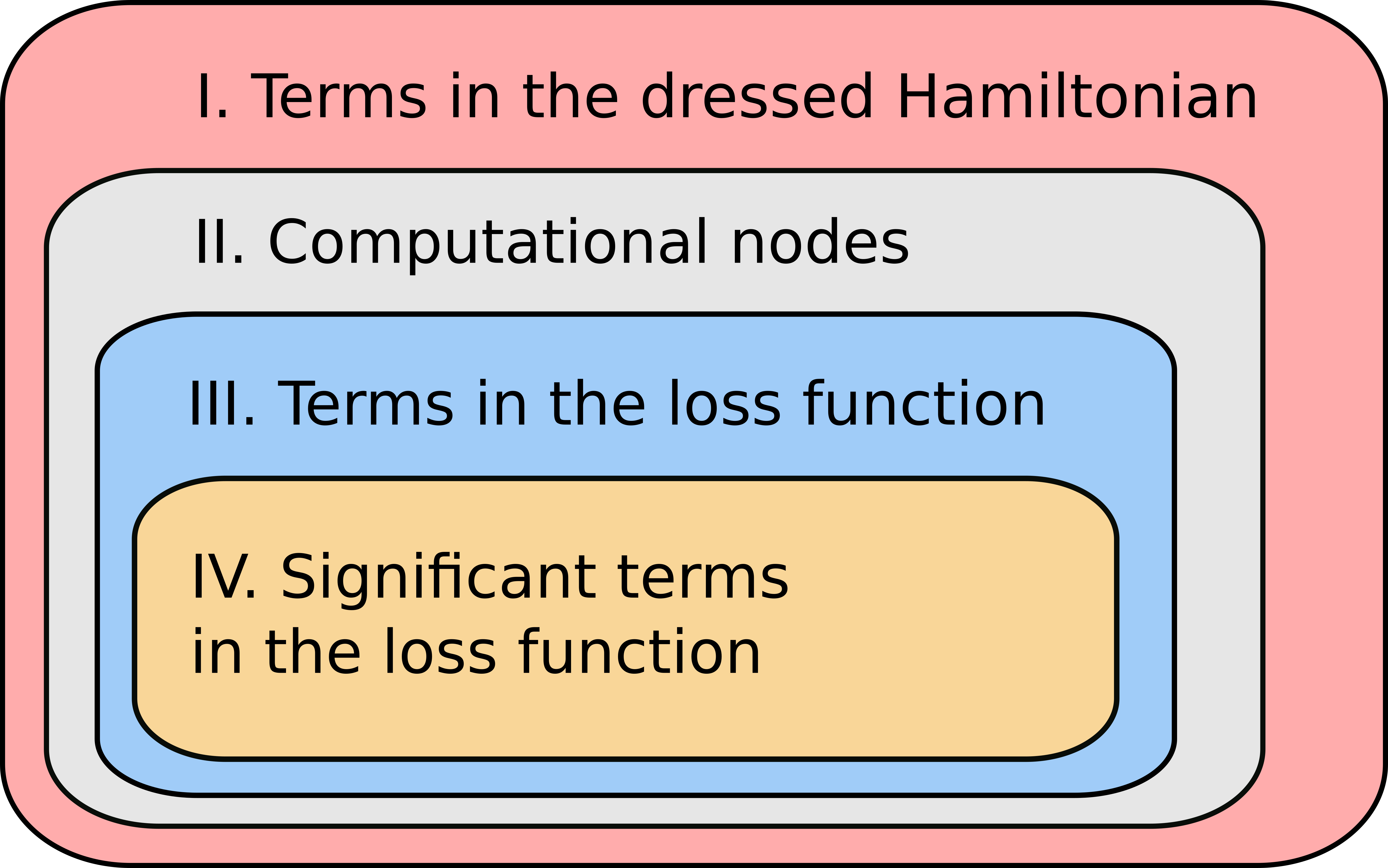} 
        \caption{}
        \label{fig scales}
        \end{subfigure}
        
        \begin{subfigure}{0.5\textwidth}
        \includegraphics[width=\hsize]{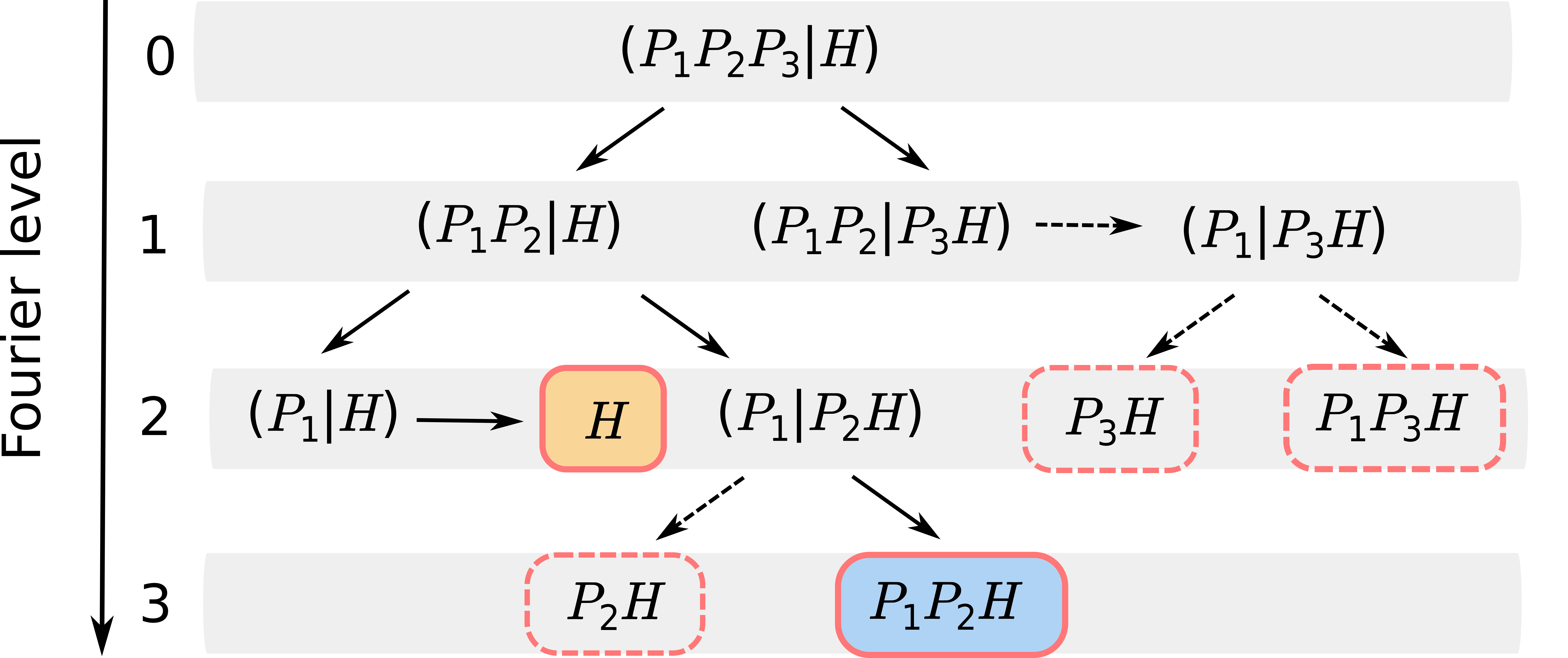}
        \caption{}
        \label{fig tree colored}
        \end{subfigure}
        
	\caption{(a) A sketch of the hierarchy of scales in the problem. (b) A sample computational tree taking into account different properties of nodes. Dashed arrows lead to nodes $(P_1\dots P_m|O)$ for which $O_X$ does not lie in the span of $(P_i)_X$. These nodes do not contribute to the loss function and are not actually generated by the algorithm. The number of remaining nodes quantifies the complexity of the algorithm. Only two final observables with non-vanishing expectation values are present, and the one at the second level contributes most to the loss function norm.}
    \label{fig complexity}
\end{figure}
As sketched in Fig.~\ref{fig scales}, our discussion features four different scales. The largest scale (I) is set by the number of terms in the expansion of the dressed Hamiltonian. It depends only on the structure of the circuit and the Hamiltonian, and can contain up to $2^M$ terms. Note that the total number of nodes in the computational tree Fig.~\ref{fig tree} can only exceed the number of final observables by a constant factor, so the recursive expansion algorithm 
is optimal for computing the Fourier series of the dressed Hamiltonian. 

Another scale (III) corresponds to the number of non-zero terms in the Fourier expansion of the loss function. It quantifies the very complexity of describing the loss function by its Fourier series, and implies a limit to when such a description can be practical. Also, as discussed in Sec.\eqref{sec truncation}, the truncated Fourier series can furnish a good approximation to the full loss function, while containing only a tiny fraction of all non-zero terms. Hence, we associate a separate scale (IV) to it.

The most relevant scale in practice, however, is set the by number of computation nodes (II). It quantifies the complexity of the algorithm. Without accounting for the expectation values, it simply coincides with the number of terms in the dressed Hamiltonian. In Sec.~\ref{sec pruning} we explained how to prune the branches of the dressed Hamiltonian expansion, with the expected saving being exponential in the number of qubits. Still, in general, this leaves a large gap between the number of computational nodes and the number of non-zero terms in the loss function. Indeed, assume for simplicity that the first $N$ Pauli generators span an $X$-basis, and consider a computational node $(P_1,\dots,P_N|O)$. For generic $P_i$ and $O$ this node is exponentially unlikely to make a non-zero contribution to the loss function. Indeed, there is a unique combination $P_1^{k_1}\dots P_N^{k_N}O$ that has a non-zero expectation value in the all-zero state, but this very combination is unlikely to satisfy all the branching rules, and hence actually appear in the recursive expansion. Hence, most nodes of the type $(P_1,\dots,P_N|O)$ do not contribute to the loss function and are not necessary to generate in the first place. Can a more efficient pruning algorithm be developed?

Let us formalize the question.  For Pauli strings $P_i, P_j$, set $\langle P_i, P_j\rangle=1$ if they anti-commute and $\langle P_i, P_j\rangle=0$ otherwise. Note that for any three Pauli strings $P_i, P_j, P_k$ it holds that $\langle{P_i,P_jP_k}\rangle=\langle P_i,P_j\rangle+\langle{P_i,P_k}\rangle$ (here and in the following  $({\rm mod}~2)$ is implied). All possible final observables in the expansion of the dressed Hamiltonian are of the form
\begin{equation}
    O(k) = P_1^{k_1}\dots P_M^{k_M}H,
\end{equation}
with some $k\in\mathbb{F}_2^M$. In order for $O(k)$ to actually appear in the set of final observables, $k$ has to be consistent with the branching rules. If $k_i=1$, the Pauli string $P_i$ must anti-commute with $P_{i+1}^{k_i+1}\dots P_M^{k_M}H$, while $k_i=0$ implies no constraints. These conditions can be expressed as $M$ equations ($i=1,\dots, M$)
\begin{align}
	k_i= k_i\braket{P_i, H}+\sum_{j=i+1}^M k_ik_j \braket{P_i, P_j} \ . \label{kikj}
\end{align}
The condition for $O(k)$ to have a non-zero expectation value is 
\begin{align}
    H_X=\sum_{i=1}^M k_i (P_i)_X=0 \ . 
    \label{kiX}
\end{align}
This relation contains $N$ constraints for an $N$-qubit problem.

Together, branching constraints \eqref{kikj} and $X$-constraints \eqref{kiX} present an instance of a boolean multivariate quadratic problem (Boolean MQ), which is known to be NP-hard. State-of-the-art algorithms \cite{Dinur2020, Barbero2022} have worst-case time complexities around $2^{0.69M}$ to find a solution or prove one does not exist. Due to the recurrent structure of equations \eqref{kikj}, our Boolean MQ instances are significantly simpler than the general case. As shown in the next section, for random circuits, which are expected to capture the worst case behavior in practice, time complexity around 
$2^{0.59M-N}$ is sufficient to find all solutions. Thus, while generic algorithms for the Boolean MQ problem are unlikely to be useful directly, there is a possibility that more efficient pruning techniques can be adopted in our scheme, narrowing the gap between the number of computational nodes and non-zero coefficients in the loss function.

\section{Case studies} \label{sec cases}
So far, we discussed general properties of the Fourier expansion for \CP{} circuits. In this section, we consider several specific examples that showcase how the expansions are structured in practice. We will both make analytic estimates and put the classical algorithm to work in numeric simulations.
\subsection{Random circuits}
\begin{figure}
	\begin{subfigure}{0.4\textwidth}
		\includegraphics[width=\hsize]{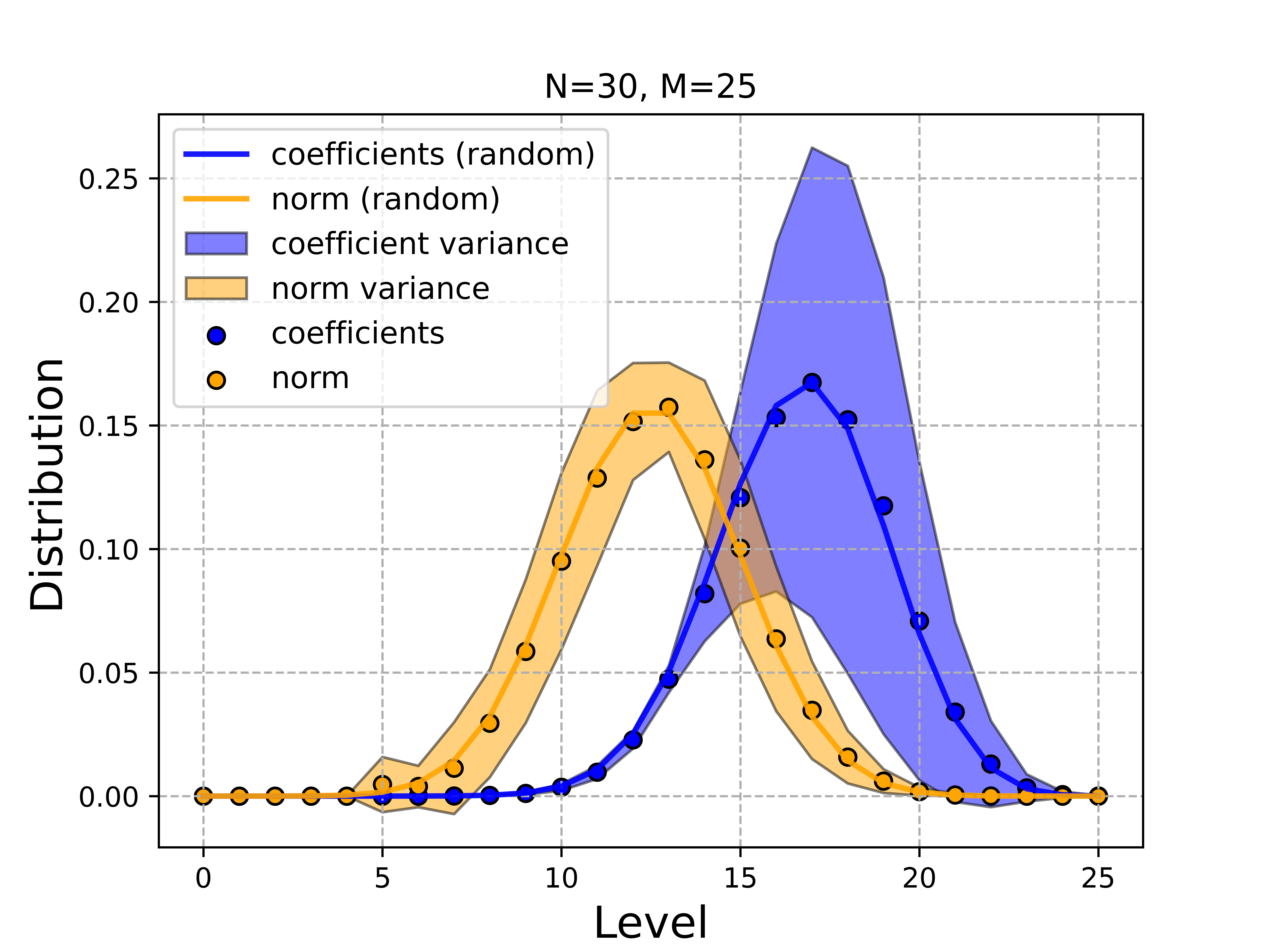}
		\caption{}
		\label{fig random nodes}
	\end{subfigure}	

	\begin{subfigure}{0.4\textwidth}
		\includegraphics[width=\hsize]{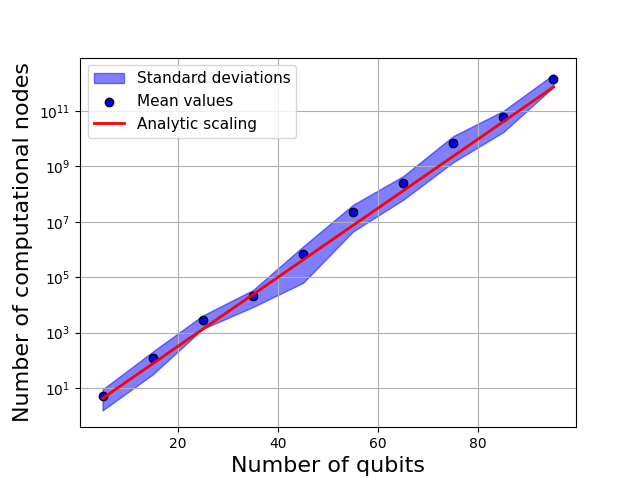}
		\caption{}		
		\label{fig random scaling}
	\end{subfigure}
	
	\caption{(a) Level distribution of the number of terms $l(m)$ and norm $\nu(m)$ (both normalized) in Fourier expansion of the dressed Hamiltonian for random circuits. Scatter plots are mean values computed from the simulations, filled areas quantify standard deviations. Solid curves are theoretical predictions. (b) Complexity of the algorithm for random Pauli circuits as a function of the number of qubits $N$ with depth $M=N/\log\frac32$. Details of numerical experiments are discussed in App.~\ref{app random}.}
	\label{fig random}
\end{figure}
We first study the case where all the Pauli generators, as well as the Hamiltonian, are random Pauli operators with the support on all of the qubits. In this setup, it is simple to give probabilistic estimates for the expected distribution of Fourier terms in the loss function. 

In fact, the assumption that the Pauli generators are random is not necessary as long as the observable is random. Therefore, we expect this behavior to capture well the asymptotic limit of most sufficiently deep circuits. Indeed, even if the original Hamiltonian is local, as we go down the computational tree (see Fig.~\ref{fig tree}), the observables at the intermediate computational nodes become ever more scrambled, and eventually behave as random Pauli operators. The argument is not rigorous, of course, and circuits that do not confirm to this pattern can be constructed. Nevertheless, the behavior of random circuits should give a useful reference point for generic circuits.\\[4pt]
\textbf{\textit{Distribution of terms in the dressed Hamiltonian}}. 

We will first look at the coarse-grained characteristics of the dressed Hamiltonian expansion, such as the number of non-zero terms at level $m$, denoted by $n_M(m)$, and the total number of terms $n_M=\sum_m n_M(m)$ (here we add a subscript $M$ to emphasize dependence on the total number of parameterized gates). When all $P_m$ and $H$ are random, the probability of branching at each computational node is $\frac12$. Therefore, on average, each iteration of the algorithm increases the total number of nodes $n_M$ by a factor of $\frac32$, leading to
\begin{align}
	n_M=\left(\frac32\right)^M \ . \label{nM}
\end{align}
The same reasoning applies to the number of non-zero computational nodes at each level $n_{M}(m)$, which hence satisfies the recurrence relation $n_{M+1}(m)=\frac12 n_{M}(m)+n_{M}(m-1)$. Solving it yields
\begin{align}
	n_{M}(m)=2^{m-M}\binom{M}{m} \ .
\end{align}
One can check that $\sum_{m=0}^M n_{M}(m)=n_M$.
\\[4pt]
\textbf{\textit{Distribution of terms in the loss function}}.
 
So far we discussed the distribution of Fourier terms in the dressed Hamiltonian, and now we turn to the loss function. Since every final observable of the dressed Hamiltonian is again a random Pauli, it has $2^{-N}$ probability of having non-zero expectation value. Therefore, the distribution of Fourier terms by level $l_M(m)$ is simply 
\begin{equation}
    l_M(m)=2^{-N}n_M(m).     
\end{equation}
The expected number of all non-zero terms in the loss function is
\begin{align}
	l_M = 2^{-N}\left(\frac32\right)^{M}\approx 2^{0.59M-N}\ . \label{lM}
\end{align}

Now, we recall that each term at level $m$ contributes exactly $2^{-m}$ to the $L^2$ norm. From the distribution of Fourier coefficients by level, we can derive the distribution of norm by level $\nu_{M}(m)=2^{-m}l_{M}(m)$, explicitly given by
\begin{align}
	\nu_{M}(m)=2^{-N-M}\binom{M}{m} \ . \label{binom}
\end{align}
Note that $\sum_{m=0}^M \nu_{M}(m)=2^{-N}$. At Fig.~\ref{fig random nodes} we plot results of numerical simulations for random circuits, which convincingly confirm our estimates. 
\\[4pt]
\textbf{\textit{Accuracy of a truncated expansion}}. 

We can now address the question of how many terms need to be included in the loss function to give a good $L^2$ approximation. Since binomial distribution \eqref{binom} is symmetric around $m=\frac{M}{2}$, including Fourier terms up to level $\frac{M}{2}$ will on average account for $50\%$ of the norm. The total number of nodes at the included levels can be estimated as $l_M(\frac{M}{2})\sim 2^{\frac{M}2-N}$. While in the large $M$ limit this is an exponentially small fraction of all terms \eqref{lM}, the number of relevant terms itself is still an exponential in $M$.
\\[4pt]
\textbf{\textit{Complexity of the algorithm and simulation limits.}}

Exponential increase in the number of relevant terms with $M$ clearly limits the depth of the circuits we can address. Importantly, the property that most final observables have zero expectation values in turn limits the number of qubits $N$ we can meaningfully simulate. While in principle, the number of qubits is only limited by the simulation cost of Clifford circuits, to yield a non-zero loss function the number of Pauli rotation gates $M$ needs to increase with $N$.

Requiring the number of non-zero terms in the loss function \eqref{lM} to stay of order $\mathcal{O}(1)$ as we increase $N$ requires to scale the depth of the circuit as $M\sim N/\log_2\frac32$. Thanks to the subroutine filtering by the expectation value, the algorithm only branches during processing of the first $M-N$ gates, leading to the number of computational nodes around $\sim \left(\frac32\right)^{M-N}\simeq 2^{0.59(M-N)}$. Therefore, the number of computational nodes generated per non-zero term in the loss function grows with the number of qubit roughly as
\begin{equation}
    \sim 2^{0.41N}\approx 10^{N/8}
\end{equation}
As reported in Fig.\ref{fig random scaling}, this scaling is well confirmed numerically.

With a computational budget to process $10^6$ nodes, which takes minutes with a basic implementation run on a laptop, the Fourier expansion of the loss function for a $50$-qubit random circuit with $85$ Pauli rotation gates can be computed exactly. With resources to process $10^{12}$ nodes, which should be feasible with an efficient implementation on a computational cluster, $100$-qubit circuits with about $170$ Pauli gates can be handled.

\subsection{QAOA}
\begin{table}
	\begin{tabular}{cllll}
		\toprule
		degree$\backslash$level&  1 & 2 & 3 & 4\\ \hline
		2  &$10^{0.7\pm0.08}$  &$10^{1.5\pm0.9}$  &$10^{3.4\pm1.7}$ & $10^{4.7\pm 2.6}$ \\
		3  &$10^{1.1\pm0.1}$  &$10^{3.6\pm0.4}$ &$10^{8.6\pm0.9}$ & $10^{17.5\pm 2.2}$  \\ 
		4  &$10^{1.4\pm0.3}$  &$10^{5.6\pm0.5}$ &$10^{16.4\pm1.3}$ & $10^{36.6\pm 3.0}$ \\
		\bottomrule
	\end{tabular}
	\caption{Estimated number of computational nodes to exactly compute Fourier expansion for QAOA circuits of varying degree and level. Statistics is collected over different graphs and observables, see App.~\ref{app QAOA} for details.}
	\label{tab QAOA}
\end{table}

Variational circuits appearing in practice are far from the random Pauli model described above. Instead, they typically only involve local gates and observables. If two local Pauli strings are supported on different subsets of qubits they necessarily commute, and hence the probability of two generic local operators anti-commuting is much smaller than $\frac12$. Although we expect the random Pauli model to describe well the large depth asymptotic, circuits of shorter depth may behave quite differently. Indeed, until the observables at computational nodes become sufficiently scrambled, the branching probability is much smaller than $\frac12$, and the complexity growth is much slower. This allows to compute Fourier expansions for circuits with much higher depths than anticipated for the random model.

As a case study, we consider instances of the Quantum Approximate Optimization Algorithm (QAOA) for the MaxCut problem on regular graphs \cite{Farhi2014}, which is the most studied approach to combinatorial optimization. Two key characteristics of a particular QAOA circuit is the degree of a graph $d$ and the number of layers $p$, see App.~\ref{app QAOA} for details. 

Due to locality, shallow instances of QAOA allow for efficient classical computation of the loss function independently of the number of qubits. To every observable one associates a reverse light cone, containing all qubits that are connected to the observable by the entangling gates. For graphs of bounded degree, the size of the reverse light cone stays constant in the large $N$ limit and classical computation of the loss function never involves simulating quantum circuits larger than that size. For 
QAOA on a graph of degree $d$ with $p$ layers the size of the reverse light cone is bounded by \cite{Farhi2014}
\begin{align}
	N_c=2\frac{(d-1)^{p+1}-1}{d-2} \ . \label{q QAOA}
\end{align}
Note that $N_c$ scales exponentially with the level $p$, so that large $p$ regime is the most difficult to simulate. 

Large $p$ also implies large number of parametric gates, which is the key limiting factor for our algorithm. At the same time, while the locality of gates is not necessary in our approach, it certainly helps. Moreover, the benefits of locality are incorporated automatically. Indeed, the Pauli generators supported outside the reverse light cone of the Hamiltonian commute with all observables at the computational nodes and are trivially eliminated. Using a basic Monte-Carlo sampling, we estimate the expected complexity of our algorithm to compute the full Fourier expansion of the loss function for several values of $d$ and $p$. Results are reported in Tab.~\ref{tab QAOA}, details of numerical simulations are specified in App.\ref{app QAOA}. Though estimates are very crude, they provide useful anchors for expected complexities.

We stress here that the special structure of QAOA circuits and observables makes it possible to handle much larger number of parameters compared to random circuits. Indeed, for a $d$-regular graph with $N$ nodes there are exactly $|E|=Nd/2$ edges, so the number of parametric gates in our simulations is given by
\begin{align}
	M=p(N_c+|E|)=pN_c\left(\frac{d}{2}+1\right) \ .
\end{align}
For instance, for $p=3, d=3$ this gives $N_c=30, M=225$, and a random circuit in this setup would require processing $\sim \left(\frac32\right)^{M-N_c}\simeq 2.18\times 10^{34}$ computational nodes, clearly an unmanageable amount. In contrast, for the actual QAOA instances, we find about $10^9$ computational nodes to be sufficient. The distribution of Fourier terms in QAOA appears to be somewhat irregular, although with clustering characteristics similar to the random circuits of equivalent complexity, see App.~\ref{app QAOA} for details.

\subsection{Hardware-efficient circuits}
One frequently studied design of variational algorithms is the hardware-efficient form \cite{Kandala2017}, where the circuit is constructed to give the maximal expressivity with limited depth, efficiently using native hardware gates. 

\begin{figure}
	\includegraphics[width=0.45\textwidth]{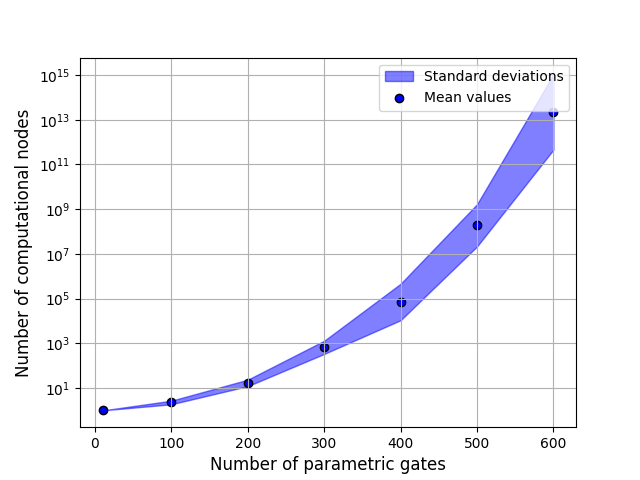}
	\caption{Estimated algorithm complexity for computing the full Fourier series of a two-local hardware efficient circuit with 50 qubits and Pauli Hamiltonian of weight two, as a function of the number of parametric gates. Details are specified in App.~\ref{app HEA}.}
	\label{fig HEA}
\end{figure}

We consider hardware-efficient circuits with brick wall architecture, where each block is built of an entangling $\CZ$ gate and four single-qubit Pauli rotations, and observables of weight $2$, see App.~\ref{app HEA} for details. In this setting, we estimate the complexity of our algorithm to exactly compute the loss function and report the results in Fig.~\ref{fig HEA}. The takeaway is similar to the QAOA case, locality of the circuit and the observable strongly reduces the number of Fourier terms compared to the random case, so the computational budget needed to process random circuits with $M\sim 180$ should be sufficient to fully characterize the Fourier series for circuits with $M\sim600$ parametric gates in the current local setup.

\section{Discussion and outlook}
We looked at some qualitative and quantitative aspects of Fourier series expansion of VQA loss functions. The main observation is that restricting to the class of \CP{} circuits allows giving a much finer picture than possible for generic VQA. We presented an efficient classical algorithm for computing the Fourier expansion level by level, with the worst case complexity bounded by ${\cal O}(N2^M)$ for a single Pauli observable. We estimated the complexity of the algorithm and characterized the distribution of Fourier terms in several examples, including random non-local \CP{} circuits as well as more conventional local circuits such as QAOA and HEA. We anticipate our findings to provide useful tools for further study of the interplay between the properties of VQA and the structure of their Fourier series expansion.

One major issue facing VQA is trainability, where two crucial obstacles are barren plateaus \cite{McClean2018, Cerezo2021a} and local minimums \cite{Bittel2021, Anschuetz2022}. Although many ansatz structures, initialization and optimization heuristics have been proposed to mitigate these problems (we refer to \cite{Wang2023} for a nice summary, and to \cite{Larocca2021, Kim2021} for the discussion of the over-parameterized setting), they persist in many practical scenarios. Interestingly, the Fourier series contains a wealth of global information about the loss landscape, and may hence give a useful perspective on these problems. For instance, we note that the variance of the loss function gradient, frequently used to diagnose barren plateaus, is naturally related to the Fourier coefficients 
\begin{align}
	\operatorname{var}_{\pmb\phi}\left(\nabla F(\pmb \phi)\right)^2=-\int d\pmb\phi\,\, F(\pmb\phi)\Delta F(\pmb\phi)=\sum_{m=0}^M m\, l(m) \ .
\end{align}
(To arrive at this expression first integrate by parts and then use the fact that for every trigonometric monomial $t_m(\pmb\phi)$ of degree $m$ it holds $\Delta t_m(\pmb\phi)=-m t_m(\pmb\phi)$.)

Therefore, already coarse-grained characteristics of the Fourier expansion, such as the distribution of terms by level $l(m)$, may provide a very useful input. Note that this distribution can be estimated by a Monte-Carlo sampling, even when the full computation of the loss function may be out of reach. Also, it would be very interesting to evaluate the role of lower vs higher order Fourier modes in shaping the loss landscape. Superficially, since the higher order modes are more oscillatory, yet typically contribute less to the $L^2$ norm of the loss function, they may be a justified suspect in creating the majority of spurious local minima.

Also, it looks promising to explore the potential of our algorithm for classical computation of quantum mean values. For circuits with local gates and constant depth, classical algorithms exist that scale favorably with the number of qubits \cite{Bravyi2021}. While gate locality helps in practice, our algorithm is based on the properties of stabilizer circuits it is not principally constrained by it. Neither it is constrained by the degree of entanglement \cite{Vidal2003, Zhou2020b} or circuit's graph treewidth \cite{Markov2005}, which can be a bottleneck for simulators based on tensor networks. Apparently, the flavor of our approach is most similar to simulation schemes for circuits dominated by Clifford gates. The parameter ranges that can be handled look similar, involving 50-100 qubits and dozens to hundreds of non-Clifford gates \cite{Bravyi2016, Bravyi2019}, but detailed benchmarks are necessary to make a reasonable comparison. We should also stress that our approach is neither a simulation algorithm (in a weak or strong sense) nor an algorithm exclusively computing mean values. Instead, it returns the full (or truncated) Fourier series of a VQA, and hence has an interesting and novel character.

\textbf{Acknowledgments}. We thank Vsevolod Yashin and V Vijendran for useful discussions. We thank the Priority 2030 program at the National University of Science and Technology “MISIS” under the project K1-2022-027.

\onecolumngrid

\appendix
\section{Structure of a generic Fourier expansion} \label{app explicit}
\subsection{Level expansion and number of terms}
For the sake of clarity, here we give a more detailed description of the general structure of Fourier expansion for the loss function introduced in Sec.~\ref{sec VQA}. Let us first illustrate trigonometric expansion of the unitary \eqref{U expansion} for the case with $M=2$ angles
\begin{align}
	U(\phi_1,\phi_2) = U_{00}\cos\frac{\phi_1}{2}\cos\frac{\phi_2}{2}+U_{01}\cos\frac{\phi_1}{2}\sin\frac{\phi_2}{2}+U_{10}\sin\frac{\phi_1}{2}\cos\frac{\phi_2}{2}+U_{11}\sin\frac{\phi_1}{2}\sin\frac{\phi_2}{2} \ . \label{U2}
\end{align}
Note that trigonometric monomials here have the same degree and period $4\pi$. Substituting such expansions into the definition of the loss function \eqref{F def} leads to the Fourier series expansion of the loss function. Applying identities $\cos^2\frac{\phi}{2}=\frac{1+\cos\phi}{2}, \sin^2\frac{\phi}{2}=\frac{1-\cos\phi}{2}, \cos\frac{\phi}{2}\sin\frac{\phi}{2}=\frac{\sin\phi}{2}$ leads to an expression involving trigonometric monomials of a smaller period $2\pi$ and degrees up to $M$. For instance, 
\begin{align}
	F(\phi_1,\phi_2)=\Braket{0|U^\dagger(\phi_1,\phi_2)HU(\phi_1,\phi_2)|0}=\frac14\Braket{0|U_{00}^\dagger HU_{00}|0}\left(1+\cos\phi_1+\cos\phi_2+\cos\phi_1\cos\phi_2\right)+\dots
\end{align}
and we wrote explicitly only the contribution from the first term.

More generally, terms in the expansion \eqref{Fm def} assume the following form
\begin{align}
F_0=\text{const},\quad
F_1(\pmb\phi)=\sum_{i=1}^M F_i(\phi_i), \quad F_2(\pmb\phi)=\sum_{i,j=1}^M F_{ij}(\phi_i,\phi_j), \quad F_3(\pmb\phi)=\sum_{i,j,k=1}^M F_{ijk}(\phi_i,\phi_j,\phi_k),\quad\dots
\end{align}
Here
\begin{align}
\begin{split}
&F_i(\phi_i)=A_i\cos\phi_i+B_i\sin\phi_i\\
&F_{ij}(\phi_i,\phi_j)=A_{ij}\cos\phi_i\cos\phi_j+B_{ij}\cos\phi_i\sin\phi_j+C_{ij}\sin\phi_i\cos\phi_j+D_{ij}\sin\phi_i\sin\phi_j\\
&\dots
\end{split}
\end{align}
At each level $m$ there are $\binom{M}{m}$ subsets of parameters, enumerating possible indices of homogeneous polynomials $F_{i_1\dots i_m}$. To define the polynomial for each parameter configuration requires specifying $2^m$ coefficients. This leads to the counting \eqref{3M coeffs} for the total number of coefficients in the Fourier series.
\subsection{Coefficients from averages}
Trivially $F_0=\Braket{F(\pmb\phi)}_{\pmb\phi}$. Higher order terms can be obtained similarly. For instance, $F_i(\phi_i)=\Braket{F(\pmb\phi)-F_0}_{\pmb\phi\neq\phi_i}$, i.e. averaging over all angles except $\phi_i$ only leaves first-order monomials involving a given angle $\phi_i$. Higher order terms can be found recursively. Note that in this prescription, to compute terms at level $m$ requires to first compute and subtract the contribution of all levels below $m$.

\section{Optimizing Pauli order} \label{app pauli}
A Pauli form of a \CP{} circuit is not unique, because adjacent Pauli rotations with commuting generators can be swapped. In principle, the number of computational nodes may be sensitive to this ordering. We briefly mention two possible optimizations along these lines.

\textit{Delayed branching.} The first strategy may be to delay branching as long as possible, by swapping the anti-commuting Pauli to the left. For illustration, consider a Pauli circuit  $(P_{1}\dots P_{N} P'_{1}\dots P'_{N}|O)$ where all Pauli generators mutually commute, and in addition 
$[P_i,O]=0$. Generators $P'_i$ may not commute with $O$. While processing operators $P_i'$ up to $2^N$ nodes may be generated. After that, assuming that $P_i$ span an $X$-basis, no new nodes will be produced. Had we started with the circuit $(P'_{1}\dots P'_{N}P_{1}\dots P_{N}|O)$ instead, which is equivalent by assumption, generators $P_i$ would be eliminated right away and no branching would be required (assuming $P'_i$ also span an $X$-basis). 

\textit{Early pruning.} 
The freedom to swap commuting Pauli generators can also be used to enforce early pruning. As an illustration, consider circuit $(P P'_1\dots P'_m|O)$, where $P$ commutes with all $P'_i$ and, moreover, $P$ is an independent generator, in the sense that $P_X$ does not lie in the span of $(P_i')_X$. If we process the circuit directly, up to $2^m$ computational nodes are generated and then tested against compatibility with the last generator $P$. However, in the equivalent setting $(P'_1\dots P'_m P|O)$ compatibility with $P$ is tested right away, and on average half of observables $O$ will be pruned leading on average to $2^{m-1}$ computation nodes. If there are $n$ generators of $P$-type, savings up to $2^{-n}$ can be expected.

Note that both optimizations rely on the presence of large streaks of commuting Pauli generators. Therefore, for the non-local random Pauli circuits very little is to be gained, while, for structured local circuits (e.g. QAOA) the savings might be substantial.

\section{Details of numerical computations}
An implementation of the algorithm and the data presented in the paper are available at GitHub repository \cite{FourierVQA}.
\subsection{Random circuits} \label{app random}
\begin{figure}
	\begin{subfigure}{0.32\textwidth}
		\includegraphics[width=\hsize]{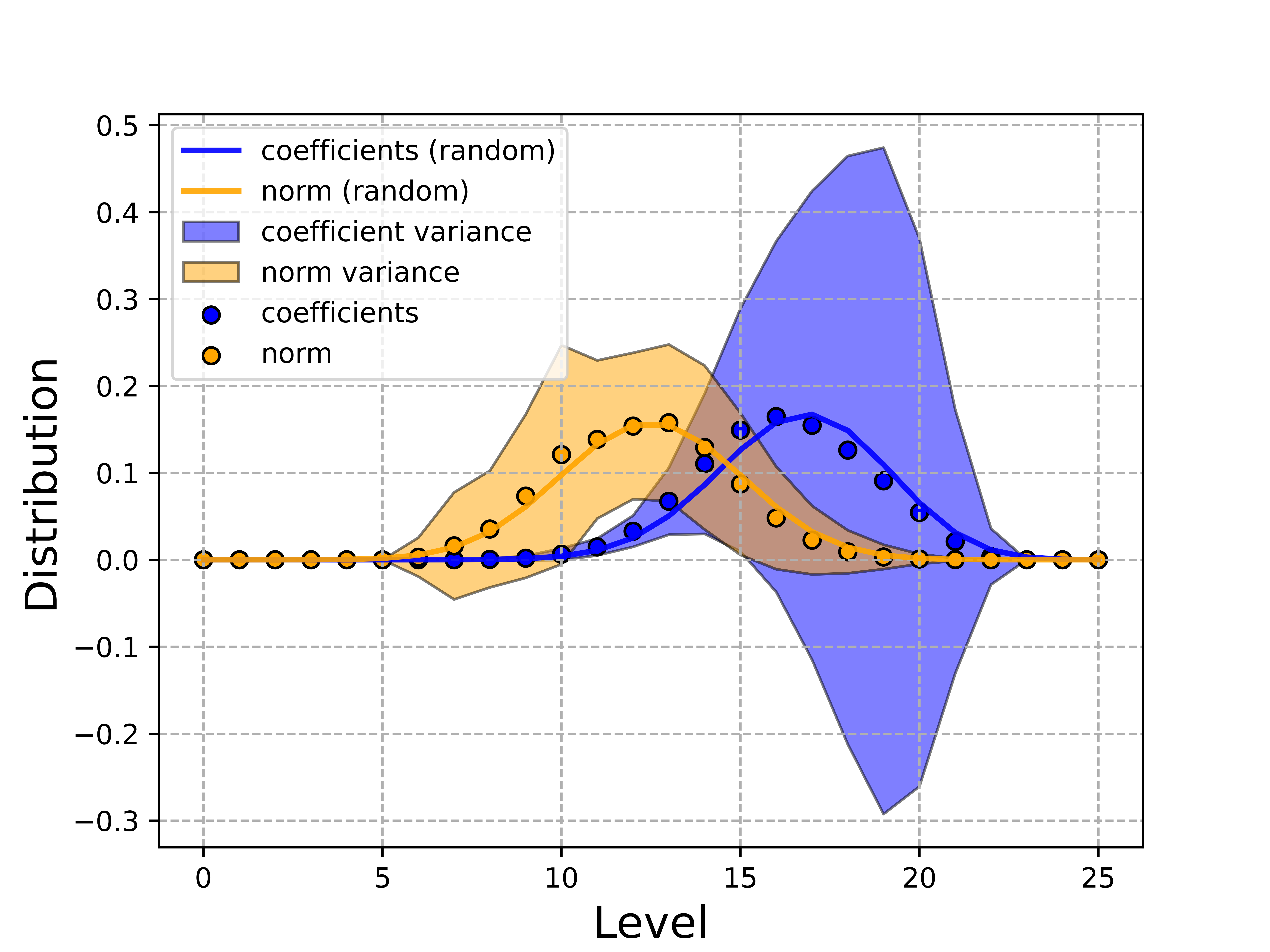}
		\caption{}
		\label{fig random local nodes}
	\end{subfigure}
	\begin{subfigure}{0.32\textwidth}
		\includegraphics[width=\hsize]{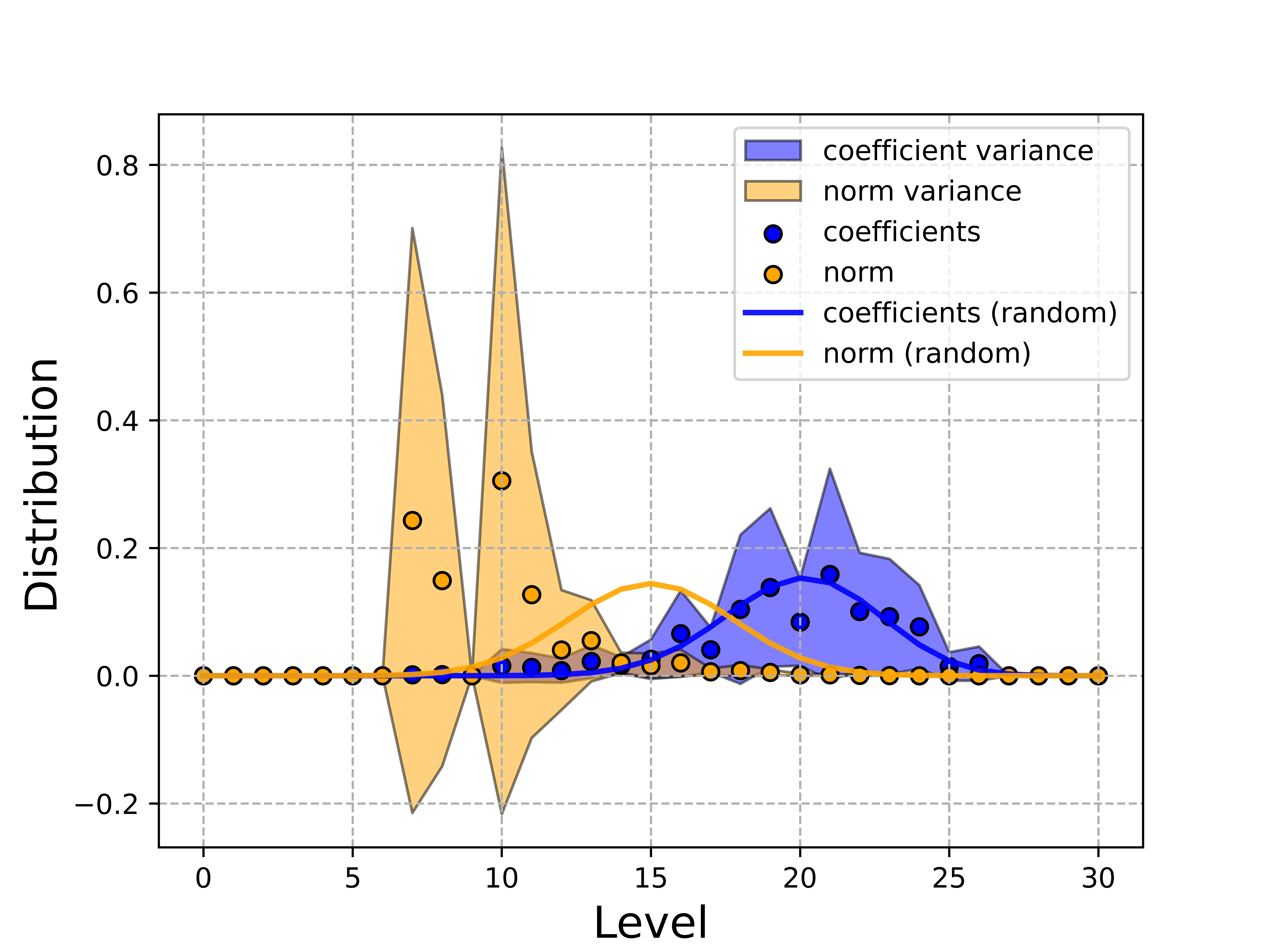}
		\caption{}
		\label{fig qaoa nodes}
	\end{subfigure}
	\begin{subfigure}{0.32\textwidth}
		\includegraphics[width=\hsize]{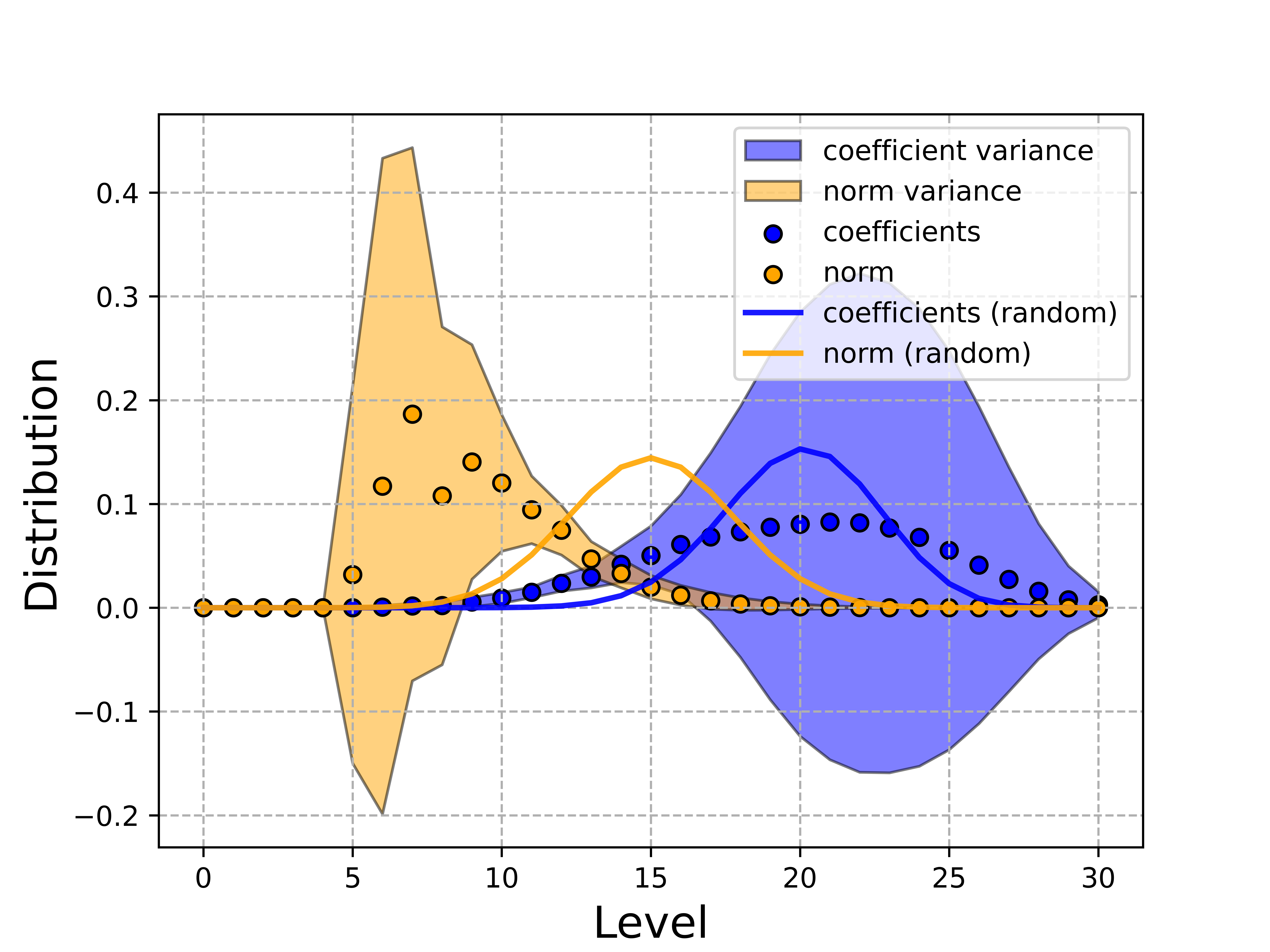}
		\caption{}
		\label{fig hea nodes}
	\end{subfigure}
\caption{Normalized distribution of non-zero Fourier coefficients in (a) local circuits with random observables (b) QAOA circuits with $d=3, p=2$ and (c) brick wall HEA circuits with $N=50$ and $M=304$. Statistics is collected for random circuits structures (except for HEA, where it is fixed) and random observables. Solid lines are theoretical curves for random circuits of the equivalent complexity.	}
\end{figure}

Statistics in Fig.~\ref{fig random nodes} is collected from 20 random non-local Pauli circuits with $N=30$ qubits and depth $M=25$. Note that the distribution of norm $\nu(m)$ is not an independent characteristic, but is computed from the distribution of nodes $l(m)$ according to $\nu(m)=2^{-m}l(m)$. In Fig.~\ref{fig random local nodes} we also plot the distribution of Fourier terms for random circuits, where only the observable is non-local, while the circuit consists of random Pauli exponentials of weight $2$. As expected, the average values are the same as for the non-local random circuits, but the fluctuations around the mean are much higher.

\subsection{QAOA} \label{app QAOA}
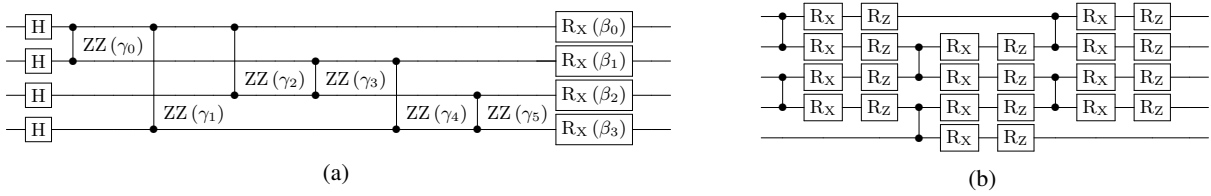
\begin{figure}
	\begin{subfigure}{0.5\textwidth}
		\scalebox{0.8}{
		\Qcircuit @C=1.0em @R=0.2em @!R { \\
			& \gate{\mathrm{H}} & \ctrl{1} & \dstick{\hspace{2.0em}\mathrm{ZZ}\,(\mathrm{\gamma_0})} \qw & \qw & \qw & \ctrl{3} & \qw & \qw & \qw & \ctrl{2} & \qw & \qw & \qw & \qw & \qw & \qw & \qw & \qw & \qw & \qw & \qw & \qw & \qw & \qw & \qw &  \gate{\mathrm{R_X}\,(\mathrm{\beta_0})} &  \qw &  \qw\\
			& \gate{\mathrm{H}} & \control \qw & \qw & \qw & \qw & \qw & \qw & \qw & \qw & \qw & \dstick{\hspace{2.0em}\mathrm{ZZ}\,(\mathrm{\gamma_2})} \qw & \qw & \qw & \ctrl{1} & \dstick{\hspace{2.0em}\mathrm{ZZ}\,(\mathrm{\gamma_3})} \qw & \qw & \qw & \ctrl{2} & \qw & \qw & \qw & \qw & \qw & \qw & \qw &  \gate{\mathrm{R_X}\,(\mathrm{\beta_1})} \qw & \qw& \qw\\
			& \gate{\mathrm{H}} & \qw & \qw & \qw & \qw & \qw & \dstick{\hspace{2.0em}\mathrm{ZZ}\,(\mathrm{\gamma_1})} \qw & \qw & \qw & \control \qw & \qw & \qw & \qw & \control \qw & \qw & \qw & \qw & \qw & \dstick{\hspace{2.0em}\mathrm{ZZ}\,(\mathrm{\gamma_4})} \qw & \qw & \qw & \ctrl{1} & \dstick{\hspace{2.0em}\mathrm{ZZ}\,(\mathrm{\gamma_5})} \qw & \qw & \qw & \gate{\mathrm{R_X}\,(\mathrm{\beta_2})} & \qw & \qw\\
			& \gate{\mathrm{H}} & \qw & \qw & \qw & \qw & \control \qw & \qw & \qw & \qw & \qw & \qw & \qw & \qw & \qw & \qw & \qw & \qw & \control \qw & \qw & \qw & \qw & \control \qw & \qw & \qw & \qw & \gate{\mathrm{R_X}\,(\mathrm{\beta_3})} & \qw & \qw}}
		\caption{}
		\label{fig QAOA circuit}
	\end{subfigure}
	\begin{subfigure}{0.45\textwidth}
		\scalebox{0.8}{
		\Qcircuit @C=1.0em @R=0.2em @!R { \\
			& \ctrl{1} & \gate{\mathrm{R_X}} & \gate{\mathrm{R_Z}} & \qw & \qw & \qw & \ctrl{1} & \gate{\mathrm{R_X}} & \gate{\mathrm{R_Z}} & \qw & \qw\\
			& \control\qw & \gate{\mathrm{R_X}} & \gate{\mathrm{R_Z}} & \ctrl{1} & \gate{\mathrm{R_X}} & \gate{\mathrm{R_Z}} & \control\qw & \gate{\mathrm{R_X}} & \gate{\mathrm{R_Z}} & \qw & \qw\\
			& \ctrl{1} & \gate{\mathrm{R_X}} & \gate{\mathrm{R_Z}} & \control\qw & \gate{\mathrm{R_X}} & \gate{\mathrm{R_Z}} & \ctrl{1} & \gate{\mathrm{R_X}} & \gate{\mathrm{R_Z}} & \qw & \qw\\
			& \control\qw & \gate{\mathrm{R_X}} & \gate{\mathrm{R_Z}} & \ctrl{1} & \gate{\mathrm{R_X}} & \gate{\mathrm{R_Z}} & \control\qw & \gate{\mathrm{R_X}} & \gate{\mathrm{R_Z}} & \qw & \qw\\
			& \qw & \qw & \qw & \control\qw & \gate{\mathrm{R_X}} & \gate{\mathrm{R_Z}} & \qw & \qw & \qw & \qw & \qw}}
		\caption{}
		\label{fig HEA circuit}
	\end{subfigure}
	\caption{Circuit layouts for (a) Single-layer QAOA  instance on a 3-regular graph with 4 nodes (b) Hardware-efficient brick wall circuit with 5 qubits (all rotation gates have different parameters, not indicated explicitly).}
\end{figure}

Given a graph $G$ with edges $E_{ij}$, a single layer of the QAOA circuit consists of two-qubit Pauli rotation gates $Z_iZ_j(\gamma_{ij})=\exp(-\imath Z_iZ_j \gamma_{ij}/2)$ (here $Z_i$ stands for a Pauli string with $Z$ on $i$-th position and identities on all others) for each edge $E_{ij}$, followed by the sequence of single-qubit $X_i(\beta_i)$ gates placed on every qubit, see Fig.~\ref{fig QAOA circuit} for an example. Note that in the standard formulation of QAOA, $\gamma_{ij}=\gamma, \beta_i=\beta$ i.e. all $ZZ$ gates and all $X$ gates have the same parameters within each layer. However, taking this correlation into account does not simplify our analysis and we will not impose it. Single layer repeated $p$ times gives an instance of QAOA with $p$ layers. The Hamiltonian is given by the sum of all $Z$-Pauli generators $H=\sum_{E_{ij}}Z_iZ_j$.

To estimate the complexity of computing the Fourier series for instances of QAOA of degree $d$ with $p$ layers, it is sufficient to consider circuits of size $N_c$ \eqref{q QAOA}. Statistics presented in Tab.~\ref{tab QAOA} was collected in the following way. For each configuration $(d, p)$, we generate 20 random regular graphs of degree $d$ and for each graph choose a single observable $Z_iZ_j$ at random. Then for each circuit-observable pair we estimate the complexity of the algorithm using the Monte-Carlo technique \ref{app MC} with $10^4$ samples. Fluctuations are significant both due to inaccuracies of the Monte-Carlo sampling, but more importantly, due to inhomogeneous nature of data gathered over different graphs and observables. To represent large fluctuations more meaningfully, we compute average values and deviations in the log scale, i.e. for the exponents of the expected number of computations nodes. Hence the format used in Tab.~\ref{tab QAOA}.  We stress that the numbers reported are estimated number of computational nodes, i.e. the expected complexity of the algorithm. The number of non-zero terms in the loss function is less by orders of magnitude, but also harder to estimate with reasonable precision.

Besides the overall complexity, it might be of interest to look at the particular distributions of the Fourier terms in the loss function. To this end, in Fig.~\ref{fig qaoa nodes}  we plot distributions on non-zero Fourier coefficients by level. Statistics is gathered for 100 different instances of $d=3,p=2$ QAOA circuits, with a single randomly chosen QAOA observable for each circuit. Theoretical curve for random circuits with $M=30$ is plotted for comparison. We note that while the distribution of the coefficients looks close enough to the random case, the distribution of norm has significant differences and large fluctuations. In particular, the norm appears to be concentrated at lower Fourier levels than expected for random circuits (after adjusting for the distribution of terms), which might make the truncated schemes quite useful.
 
\subsection{Hardware-efficient circuits} \label{app HEA}
An example of a brick wall hardware-efficient circuit that we consider is shown in Fig.~\ref{fig HEA circuit}. Statistics for Fig.\ref{fig HEA} is collected for $N=50$ qubit circuits with up to $M=600$ rotation gates. For each $M$ we take 10 random observables of weight $2$ and estimate the number of computational nodes with $10^4$ Monte-Carlo samples. Similarly to the QAOA case, due to large individual fluctuations, we compute means and deviations in the logarithmic scale.

In Fig.~\ref{fig hea nodes} we also plot the distribution of Fourier coefficients for $N=50$ circuit with $M=308$ parametric gates (corresponding to 77 \CZ gates), averaged over 100 random Pauli observables of weight 2. Similarly to the QAOA case, the most interesting part is the norm distribution, which shows large fluctuations and also the tendency to cluster at lower levels (compared to a random circuit with a similar distribution of coefficients).

\subsection{Estimating complexity from Monte-Carlo sampling} \label{app MC}
It is useful to estimate the runtime/complexity of the algorithm without performing the full computation. To do this, we can use a simple version of Monte-Carlo sampling to probe the structure of the computational tree Fig.~\ref{fig tree}. First, we will describe a version that allows to estimate the number of terms in the expansion of the dressed Hamiltonian. To this end, we can use the basic algorithm to traverse the computational tree, but instead of keeping both branches at each split, we choose one at random and disregard the other. This procedure produces a single complete branch from the computational tree. The probability to generate any particular branch is $2^{-m}$ with $m$ being the number of splits performed in the process. Let $s(m)$ be the number of samples at with $m$ splits. We estimate the total number of terms $n$ as
\begin{align}
	n\approx \frac{\sum_{m=0}^M 2^ms(m)}{\sum_{m=0}^M s(m)} \ . \label{n sampled}
\end{align}
Note that in this case the number of splits $m$ is the same as the Fourier level of the resulting term.

To estimate the complexity of the actual algorithm, i.e. the number of computation nodes generated during the computation, we need to take into account the pruning based on the expectation values, see Sec.~\ref{sec pruning}. The sampling prescription above is modified in a simple way. We traverse the computational tree and choose branches at random. Importantly, we only choose from the branches compatible with the pruning restrictions. If one of the branches in not admissible, the actual algorithm will not generate additional nodes and the probability of the sample does not need to be updated. In this case, the Fourier level of the resulting term does not correspond to the probability of sampling. The prescription \eqref{n sampled} remains unchanged.

\bibliographystyle{apsrev4-1}
\bibliography{library.bib, additional_library.bib}

\end{document}